\documentclass[conference,letterpaper]{IEEEtran}

\usepackage[utf8]{inputenc} 
\usepackage[T1]{fontenc}
\usepackage{url}
\usepackage{ifthen}
\usepackage{cite}
\usepackage[cmex10]{amsmath} % Use the [cmex10] option to ensure complicance
\usepackage{algorithmic}
\usepackage{algorithm}
\usepackage{graphicx}
\usepackage{cite}
\usepackage{array}
\usepackage{amsmath,amsfonts}
\usepackage{amssymb}
\usepackage{microtype}
\usepackage{comment}

\newtheorem{theorem}{Theorem}
\newtheorem{definition}{Definition}   % 定义
\newtheorem{lemma}{Lemma}      % 引理
\newtheorem{corollary}{Corollary}    % 推论
\newtheorem{proposition}{Proposition}  % 命题
\newtheorem{remark}{Remark}

\def\BibTeX{{\rm B\kern-.05em{\sc i\kern-.025em b}\kern-.08em
    T\kern-.1667em\lower.7ex\hbox{E}\kern-.125emX}}

\IEEEoverridecommandlockouts
\begin{document}
\title{ Counting and Entropy Bounds for Structure-Avoiding Spatially-Coupled LDPC Constructions 
\thanks{This work is partially supported by the National Key R\&D Program of China, (2023YFA1009600).}}

 \author{
 \IEEEauthorblockN{Lei Huang}
 \IEEEauthorblockA{Data Science Institute\\
                    Shandong University\\
                    Jinan, China\\
                    Email: leihuang@mail.sdu.edu.cn}}

\maketitle

\begin{abstract}
Designing large coupling memory quasi-cyclic spatially-coupled LDPC (QC-SC-LDPC) codes with low error floors requires eliminating specific harmful substructures (e.g., short cycles) induced by edge spreading and lifting. Building on our work~\cite{r15} that introduced a Clique Lovász Local Lemma (CLLL)-based design principle and a Moser--Tardos (MT)-type constructive approach, this work quantifies the size and structure of the feasible design space. Using the quantitative CLLL, we derive explicit lower bounds on the number of feasible edge-spreading and lifting assignments satisfying a given family of structure-avoidance constraints, and further obtain bounds on the number of non-equivalent solutions under row/column permutations. Moreover, via R\'enyi entropy bounds for the MT distribution, we provide a computable lower bound on the number of distinct solutions that the MT algorithm can output, giving a concrete diversity guarantee for randomized constructions. Specializations for eliminating 4-cycles yield closed-form bounds as functions of system parameters, offering a principled way to select the memory and lifting degree and to estimate the remaining search space.
\end{abstract}

\section{Introduction}
The threshold-saturation effect of spatially-coupled low-density parity-check (SC-LDPC) codes \cite{r2,r3,r4} has fueled extensive interest in these coupled ensembles. This phenomenon, together with their capacity-achieving behavior on general binary memoryless channels \cite{r5}, positions SC-LDPC codes as a powerful family of modern channel codes. Practically, the inherent coupled regularity further enables windowed decoding (WD) with low decoding latency \cite{r6,r7}. Spatial coupling traces back to the convolutional LDPC construction of Felström and Zigangirov \cite{r1}, and analogous coupled designs have recently begun to attract attention in the quantum setting (quantum SC-LDPC codes) \cite{r8,r9,r10}.

Industrial deployments place SC-LDPC codes at markedly different operating points and under heterogeneous system constraints. Power-limited links (e.g., wireless and satellite) often operate close to the iterative decoding threshold, where the waterfall behavior and rate/latency tradeoffs dominate, whereas data-storage and high-speed optical/Ethernet interfaces typically demand extremely low residual error probabilities, making the error-floor behavior a primary bottleneck. Consequently, application-driven SC-LDPC code design is inherently multi-objective: it requires both threshold-oriented optimization (e.g., via density evolution) and finite-length graph conditioning to mitigate dominant harmful substructures such as short cycles and trapping/absorbing sets. This multi-objective viewpoint has been considered, e.g., in \cite{r11}. Motivated by these diverse requirements, it becomes essential to quantify the feasible design space; in particular, establishing a lower bound on the number (or fraction) of SC-LDPC code realizations that satisfy prescribed error-floor constraints is crucial for assessing design feasibility and guiding practical code search.

Algebraic constructions can deterministically control girth and certain substructures, but they are typically restricted to narrow templates and thus do not readily yield explicit lower bounds on how many SC-LDPC code realizations satisfy prescribed structure-avoidance constraints. In contrast, randomized/heuristic design explores a much larger space—e.g., the optimal overlap-circulant power optimization (OO-CPO) framework \cite{r12} (extended in \cite{r13}) and the probabilistic optimization approach \cite{r14}—yet its search runtime can be highly instance-dependent; our Moser-Tardos (MT) resampling approach in \cite{r15} can be viewed as a randomized OO-CPO special case that locally resamples partitioning/lifting variables whenever a constraint is violated.

In this paper, we extend our existence and constructive results in~\cite{r15} from feasibility to quantitative guarantees on the solution-space size, by deriving explicit counting and entropy bounds for structure-avoiding SC-LDPC code constructions. Due to space constraints, full proofs and detailed discussions are deferred to~\cite{r19}.

\noindent\textbf{Contributions:}
(i) Under joint non-uniform edge spreading and non-uniform lifting, we apply a quantitative clique LLL to derive explicit lower bounds on the number of structure-avoiding QC-SC-LDPC designs, including bounds on non-equivalent solutions under row/column permutations;
(ii) we lower bound the number of distinct solutions producible by the MT algorithm via Rényi entropy bounds of the MT output distribution, with closed-form specializations for eliminating $4$-cycles;
(iii) we connect the MT resampling construction to the OO-CPO design framework by proving that circulant power optimization (CPO) subsumes MT as a special case (Proposition~\ref{p2}); consequently, the output-diversity guarantee in (ii) provides theoretical evidence for the strong potential of CPO-based QC-SC-LDPC design procedures  \cite{r12,r13,r14}.

\section{PRELIMINARIES}

\subsection{QC-SC-LDPC Codes}
We consider Type-I protograph-based QC-SC-LDPC codes constructed from a QC-LDPC block code. The QC structure enables efficient encoder/decoder implementations  and can endow the resulting coupled ensembles with anytime-reliable properties.

Starting from an all-one $(\gamma,\kappa)$ base matrix $\mathbf{H}$ (a fully-connected bipartite base graph), spatial coupling is realized via \emph{edge spreading}: we partition the $\gamma\kappa$ edges into $m+1$ component matrices $\{\mathbf{H}_0,\ldots,\mathbf{H}_m\}$ such that
$\mathbf{H}=\sum_{\ell=0}^{m}\mathbf{H}_\ell$ and each $1$-entry of $\mathbf{H}$ appears in exactly one $\mathbf{H}_\ell$.
The spreading is recorded by the \emph{partition matrix} $\mathbf{P}$, where for $i\in[\gamma]$, $j\in[\kappa]$, we set $\mathbf{P}(i,j)=\ell$ if and only if $\mathbf{H}_\ell(i,j)=1$ (and $\mathbf{H}_{\ell'}(i,j)=0$ for all $\ell'\neq \ell$).
For implementation-oriented designs, we assume all uncoupled sections share the same base matrix and the same edge-spreading pattern.
Stacking $\mathbf{H}_0,\ldots,\mathbf{H}_m$ vertically forms a \emph{replica}; concatenating $L$ identical replicas horizontally yields the coupled protograph matrix $\mathbf{H}^{P}_{SC}$, where $L$ is the \emph{coupling length}.

Given $\mathbf{H}^{P}_{SC}$, we obtain a QC-SC-LDPC parity-check matrix via $Z$-lifting.
Let $\mathbf{L}$ be a $\gamma\times\kappa$ \emph{lifting matrix}, and let $\sigma$ denote the $Z\times Z$ circulant permutation matrix obtained by a one-step left cyclic shift of the identity.
For each $i\in[\gamma]$, $j\in[\kappa]$, if $\mathbf{L}(i,j)=x\in\{0,\ldots,Z-1\}$, then every $1$ in $\mathbf{H}^{P}_{SC}$ corresponding to the base edge $(i,j)$ is replaced by $\sigma^{x}$, and every $0$ is replaced by the $Z\times Z$ all-zero matrix.
Under this notation, a Type-I QC-SC-LDPC code is specified by $(\gamma,\kappa,m,Z,L)$ together with $(\mathbf{P},\mathbf{L})$.

\subsection{Major Harmful Structures}
Although spatial coupling improves iterative decoding thresholds, QC-SC-LDPC codes may still exhibit error floors at high SNR over the AWGN channel, primarily due to small subgraphs (e.g., trapping/absorbing structures) that are ultimately built upon short cycles. We summarize the key cycle-activation conditions.

\begin{lemma}\label{lem_1}
Let $\mathcal{C}_{2g}$ denote the set of length-$2g$ cycle candidates in the $(\gamma,\kappa)$ base matrix, where $g\in\mathbb{N}$ and $g\ge 2$.
For $c_{2g}\in\mathcal{C}_{2g}$ represented by $(j_1,i_1,j_2,i_2,\ldots,j_g,i_g)$ (with $j_{g+1}=j_1$), the nodes $(i_k,j_k)$ and $(i_k,j_{k+1})$ for $1\le k\le g$ correspond to entries in $\mathbf{H}$, $\mathbf{P}$, and $\mathbf{L}$.
Then $c_{2g}$ remains a cycle candidate in the coupled protograph after edge spreading if and only if \cite{r16}
\begin{align}\label{e_1}
\sum_{k=1}^{g} \mathbf{P}(i_k, j_k) \;=\; \sum_{k=1}^{g} \mathbf{P}(i_k, j_{k+1}) .
\end{align}
Moreover, such a protograph cycle becomes an actual cycle in the Tanner graph after $Z$-lifting if and only if \cite{r17}
\begin{align}\label{e_2}
 \sum_{k=1}^{g} \mathbf{L}(i_k, j_k) \;\equiv\; \sum_{k=1}^{g} \mathbf{L}(i_k, j_{k+1}) \pmod Z .
\end{align}
\end{lemma}

This naturally motivates the common two-stage design strategy: first choose $\mathbf{P}$ to suppress protograph cycle candidates, then select $\mathbf{L}$ to further eliminate cycles (and induced harmful substructures) after lifting, as in \cite{r12,r14}.
In this work, we focus on the AWGN setting, where dominant error-floor mechanisms are often attributed to low-weight absorbing sets (ASs)~\cite{r12}. Moreover, all harmful structures considered in this paper are \emph{avoidable} in the sense that they can be excluded by imposing explicit design constraints within our CSP framework.

\begin{definition}[Absorbing Sets]\label{def_1}
Let $\mathcal{V}$ be a set of $a$ variable nodes (VNs) in a Tanner graph, and let $\mathcal{N}(\mathcal{V})$ be its neighboring check nodes (CNs).
Partition $\mathcal{N}(\mathcal{V})=\mathcal{N}_o(\mathcal{V})\cup \mathcal{N}_e(\mathcal{V})$, where $\mathcal{N}_o(\mathcal{V})$ (resp.\ $\mathcal{N}_e(\mathcal{V})$) denotes the CNs having odd (resp.\ even) degree within the subgraph induced by $\mathcal{V}$.
Then $\mathcal{V}$ is an $(a,b)$ absorbing set if $|\mathcal{N}_o(\mathcal{V})|=b$ and every VN in $\mathcal{V}$ has strictly more neighbors in $\mathcal{N}_e(\mathcal{V})$ than in $\mathcal{N}_o(\mathcal{V})$.
\end{definition}

\subsection{Constraint Satisfaction Problem}
In a \textit{constraint satisfaction problem} (CSP), there is a set of variables and a set of constraints imposed on these variables. The task is to find an assignment of values to the variables such that all constraints are satisfied. By limiting the types of relations used to define constraints, certain constrained versions of the CSP are known to align with many fundamental problems, such as SAT, graph coloring, and the solvability of systems of linear equations.

Disregarding the types of constraints, the dependency structures between variables and constraints can be depicted using hypergraphs or bipartite graphs. A powerful model is the \textit{variable-generated system} of events, where each event is a constraint imposed on a set of continuous or discrete independent random variables. Assume a set of events $\mathcal{A}=\{A_1,\dots,A_n\}$ and a set of variables $\chi=\{X_1,\dots,X_{m'}\}$. For each $i \in [n]$, let $\chi_i \subseteq \chi$ be the set of variables that completely determine the event $A_i$. Represent the events and variables as two sets of nodes in a bipartite graph, where an edge $(i,j) \in [n] \times [m']$  exists if and only if $X_j \in \chi_i$, resulting in the \textit{event-variable graph} $G_E=([n],[m'],E)$.

\begin{proposition}[QC-SC-LDPC design as a CSP]\label{pro_1}
Fix $(\gamma,\kappa,m,Z)$ and a finite collection of harmful structures
$\mathcal{H}=\{H_1,\ldots,H_n\}$ in the $\gamma\times\kappa$ base graph
(e.g., $\mathcal{H}=\mathcal{C}_{2g}$).
Let the variable set be
\[
\begin{aligned}
\chi &:= \{X_{i,j} : i\in[\gamma],\, j\in[\kappa]\},\\
X_{i,j} &:= (\mathbf{P}(i,j),\mathbf{L}(i,j)) \in \{0,\ldots,m\}\times \mathbb{Z}_Z .
\end{aligned}
\]

For each $H_\ell\in\mathcal{H}$, define a constraint (bad event) $A_\ell$
that $H_\ell$ becomes \emph{active} after edge spreading and $Z$-lifting.
Then finding a feasible pair $(\mathbf{P},\mathbf{L})$ that avoids all $H_\ell$ is exactly the CSP
\[
\text{find an assignment of }\chi \text{ such that } \bigcap_{\ell=1}^n \overline{A_\ell}\text{ holds.}
\]
Moreover, the corresponding event-variable graph $G_E$ is obtained by connecting
each event node $A_\ell$ to the variable nodes $\{X_{i,j}\}$ involved in $H_\ell$.
\end{proposition}

\begin{IEEEproof}
A CSP is specified by (i) a set of variables and (ii) a set of constraints on these variables.
Here the variables are the $\gamma\kappa$ edge-variables $X_{i,j}=(\mathbf{P}(i,j),\mathbf{L}(i,j))$.
Each harmful structure $H_\ell$ induces a local “activation” condition after edge spreading and lifting,
so we impose the constraint $\overline{A_\ell}$ that forbids this activation.
Importantly, whether $A_\ell$ occurs depends only on the entries of $(\mathbf{P},\mathbf{L})$ on the edges participating in $H_\ell$
(for cycle candidates this is exactly the local conditions (1) and (2) in Lemma~1).
Hence each constraint $\overline{A_\ell}$ involves only a subset of variables, and the feasibility problem
\(\bigcap_{\ell=1}^n \overline{A_\ell}\) is precisely a CSP. The bipartite graph $G_E$ is then formed by linking
$A_\ell$ to the variables it depends on, as in Sec.~II-C.
\end{IEEEproof}
Building on our earlier work \cite{r15}, we continue to formulate structure-avoiding QC-SC-LDPC code design as a CSP and employ CLLL--based tools to characterize ensembles relevant to the high-SNR (error-floor) regime. The method for utilizing CLLL-based tools within the CSP framework is detailed in Appendix F of \cite{r19}. In contrast to \cite{r15}, we further quantify the feasible design space via counting and entropy bounds.

\section{ Lower Bound on the Number of Solutions}
\subsection{Lower Bound on the Number of Solutions}

Under the CSP formulation in Proposition~\ref{pro_1}, each QC-SC-LDPC design corresponds to an assignment of the $\gamma\kappa$ edge variables
$X_{i,j}=(\mathbf{P}(i,j),\mathbf{L}(i,j))$, i.e., a pair $(\mathbf{P},\mathbf{L})$ consisting of an edge-spreading (partition) matrix $\mathbf{P}$ and a lifting-shift matrix $\mathbf{L}$.
Therefore, using the quantitative version of the CLLL, we can not only obtain existence results but also establish an explicit lower bound on the
number of feasible pairs $(\mathbf{P},\mathbf{L})$ that satisfy all structure-avoidance constraints, thereby quantifying the size of the feasible solution space.
The size and structural information of this solution space are crucial for finding codes with good overall performance at lower complexity.
In this subsection, we first use the quantitative CLLL to derive a theoretical lower bound on the number of feasible solutions $(\mathbf{P},\mathbf{L})$.

\begin{theorem}\label{the_1}
For any QC-SC-LDPC code constructed with respect to a $\gamma \times \kappa$ fully-connected block code base graph, with no parallel edges,
under the coupling pattern $\mathbf{a}=(a_0,a_1,\ldots,a_{m_t})$ and corresponding (non-uniform) distribution
$\mathbf{p}=(p_0,p_1,\ldots,p_{m_t})$,
and under the lifting-shift alphabet $\mathbf{b}=(b_0,b_1,\ldots,b_{Z-1})=(0,1,\ldots,Z-1)$ with distribution
$\mathbf{q}=(q_0,q_1,\ldots,q_{Z-1})$. Assume the random construction chooses each $\mathbf{P}(i,j)$ independently with $\Pr[\mathbf{P}(i,j)=\ell]=p_\ell$,
and chooses each lifting shift $\mathbf{L}(i,j)$ independently with $\Pr[\mathbf{L}(i,j)=b_t]=q_t$,
and $P$ is independent of $L$.

Define
\[
p_{\max}:=\max_{0\le \ell\le m_t} p_\ell,\qquad
q_{\max}:=\max_{t\in\mathbb{Z}_Z} q_t .
\]
Let $\mathcal{H}=\{H_1,\ldots,H_k\}$ be a family of avoidable harmful structures with dependency graph $G$,
and let $\{P_{H_1},\ldots,P_{H_k}\}$ be the probabilities that they are active under the above random edge-spreading and lifting process.
If
\begin{align}
    \max\{P_{H_1},\ldots,P_{H_k}\} \le \max\{\underbrace{\frac{(\Delta-1)^{\Delta-1}}{(\Delta)^{\Delta}}}_{I},\underbrace{\frac{(|H| - 1)^{|H| - 1}}{(W_\mathcal{H} - 1)|H|^{|H|}}}_{II} \}
    \nonumber 
\end{align}
where $\Delta$ is the maximum degree of the dependency graph and $|H|=\max\{|H_1|,\ldots,|H_k|\}$,
then
\begin{equation}
\begin{aligned}
\#\!\left(\bigcap_{i\in[k]}\overline{H_i}\right) \ge {} &
\left(\frac{1}{p_{\max}q_{\max}}\right)^{\gamma \kappa} \\
&\times
\begin{cases}
\left(1-\frac{2}{\Delta}\right)^{||G||}, & \text{if } I>II,\\[2pt]
\left(1-\frac{W_{\mathcal H}}{|H|(W_{\mathcal H}-1)}\right)^{||G||}, & \text{otherwise.}
\end{cases}
\end{aligned}
\end{equation}

Here, $\#(\bigcap_{i\in [k]}\Bar{H_i})$ denotes the number of feasible assignments of $(\mathbf{P},\mathbf{L})$ (equivalently, solutions in the CSP variable set $\chi$)
that make $\bigcap_{i\in[k]}\Bar{H_i}$ true.
\end{theorem}

\begin{IEEEproof}
See Appendix A of ~\cite{r19}, the notation is consistent with that in ~\cite{r15}.
\end{IEEEproof}

\begin{corollary}[uniform spreading and uniform lifting]\label{cor:uniform_C4}
For any QC-SC-LDPC code constructed with respect to a $\gamma \times \kappa$ fully-connected block code base graph, with no parallel edges,
assume uniform edge-spreading and uniform lifting,
If
\begin{equation}
    \frac{2m^2+4m+3}{3(m+1)^3Z} \le \text{Max}\{\underbrace{\frac{(\Delta-1)^{\Delta-1}}{(\Delta)^{\Delta}}}_{I},\underbrace{\frac{27}{256(\gamma \kappa -\gamma -\kappa)}}_{II} \}
\end{equation}
where $\Delta=(2\gamma-3)(2\kappa-3)$, then
\begin{align}
{\#\Big(\bigcap_{c_4\in \mathcal{C}_4}\Bar{c_4}\Big)} \ge &
[Z(m_t+1)]^{\gamma \kappa} \\
& \times
\begin{cases}
(1-\frac{2}{\Delta})^{\binom{\gamma}{2}\binom{\kappa}{2}},&{\text{if}}\ I>II, \\[2pt]
\Big(1-\frac{\gamma \kappa-\gamma-\kappa+1}{4(\gamma \kappa-\gamma-\kappa)}\Big)^{\gamma\kappa},&{\text{otherwise.}} 
\end{cases}
\end{align}
\end{corollary}

\begin{IEEEproof}
This is a specialization of the $(\mathbf p,\mathbf q)$-version corollary by setting
$p_{\max}=\frac{1}{m_t+1}$ and $q_{\max}=\frac{1}{Z}$, so that
$\big(\frac{1}{p_{\max}q_{\max}}\big)^{\gamma\kappa}=[Z(m_t+1)]^{\gamma\kappa}$.
Substituting these into the general bound gives the result.
\end{IEEEproof}
It is straightforward to verify that the expression above takes the form $\exp\big(\Theta(\gamma\kappa)\big)$, implying that the solution space is exponentially large.

\begin{definition}[column and row permutation]
 A column permutation of matrix $ M$ with $ \kappa $ columns is denoted by a vector $ \pi_c = [\rho_1, \ldots, \rho_{\kappa}]$, that is a permutation of the elements in the vector $ [1, \ldots, \kappa] $. When $ M \overset{\pi_c}{\rightarrow} M' $, $ A' $ is obtained from $ M $ such that the $ j $-th column of $ M $ is the $ \rho_j $-th column of $ M' $.  

Similarly, a row permutation of matrix $ M $ with $ \gamma $ rows is denoted by a vector $ \pi_r = [\nu_1, \ldots, \nu_{\gamma}] $, that is a permutation of the elements in the vector $ [1, \ldots, \gamma] $. When $ M \overset{\pi_r}{\rightarrow} M' $, $ M' $ is obtained from $ M $ such that the $ i $-th row of $M $ is the $ \nu_i $-th row of $ M' $.   
\end{definition}

\begin{definition}[equivalent matrices]
Two binary matrices $ M $ and $ M' $ are column-wise equivalent (resp., row-wise equivalent) if they can be obtained by column (resp., row) permutations of each other. Two binary matrices $ M $ and $ M' $ are equivalent if they can be derived from each other by a sequence of row and column permutations.    
\end{definition}

 \begin{remark}
In this paper, we employ the concept of equivalence to demonstrate that the decoding threshold and the population of combinatorial objects (such as cycles, absorbing sets \cite{r20}, and trapping sets \cite{r21}) remain invariant under row and column permutations of a binary matrix.
 \end{remark}

Lemma \ref{rc} demonstrates the structural congruence between coupled proto-matrices and their constituent partitioning matrices. Consequently, the design of coupled SC-LDPC codes can be effectively reduced to searching within a compact space of small $(\gamma \times \kappa)$ non-equivalent partitioning matrices.

\begin{lemma}{(\cite{r11})}\label{rc}
 Any column/row permutation of the partitioning matrix $\textbf{P}$ of an SC code results in an SC proto-matrix that is a column/row permuted version of the original SC proto-matrix $H^P_{\text{SC}}$ .   
\end{lemma}

\begin{theorem}
For any QC-SC-LDPC code constructed with respect to a $\gamma \times \kappa$ fully-connected block code base graph, with no parallel edges, under the coupling pattern $\mathbf{a}=(a_0,a_1,...,a_{m_t})$ and corresponding probability distribution $\mathbf{p}=(p_0,p_1,...,p_{m_t})$. If $\{H_1,...,H_k\}$ are a series of avoidable harmful structures in the base graph with dependency graph $G$, and $\{P_{H_1},...,P_{H_k}\}$ are the probabilities that they are active after the random edge-spreading and lifting process,  if
\begin{align}
    \max\{P_{H_1},\ldots,P_{H_k}\} \le \max\{\underbrace{\frac{(\Delta-1)^{\Delta-1}}{(\Delta)^{\Delta}}}_{I},\underbrace{\frac{(|H| - 1)^{|H| - 1}}{(W_\mathcal{H} - 1)|H|^{|H|}}}_{II} \}
    \nonumber 
\end{align}
then
\begin{equation}
\begin{aligned}
\#_N\left(\bigcap_{i\in[k]}\overline{H_i}\right) \ge {} &
\left(\frac{1}{p_{\max}q_{\max}}\right)^{\gamma \kappa} \left(\frac{1}{\gamma !\kappa !}\right)\\
&\times
\begin{cases}
\left(1-\frac{2}{\Delta}\right)^{||G||}, & \text{if } I>II,\\[2pt]
\left(1-\frac{W_{\mathcal H}}{|H|(W_{\mathcal H}-1)}\right)^{||G||}, & \text{otherwise.}
\end{cases}
\end{aligned}
\end{equation}
Where $\#_{N}(\bigcap_{i\in [k]}\Bar{H_i})$ is the number of all non-equivalent partition matrices that make $\bigcap_{i\in [k]}\Bar{H_i}$ true.    
\end{theorem}

\begin{corollary}
 For any QC-SC-LDPC code constructed with respect to a $\gamma \times \kappa$ fully-connected block code base graph, with no parallel edges, if
\begin{equation}
    \frac{2m^2+4m+3}{3(m+1)^3Z} \le \text{Max}\{\underbrace{\frac{(\Delta-1)^{\Delta-1}}{(\Delta)^{\Delta}}}_{I},\underbrace{\frac{27}{256(\gamma \kappa -\gamma -\kappa)}}_{II} \}
\end{equation}
where $\Delta=(2\gamma-3)(2\kappa-3)$ ,$Z$ is the lifting degree.
then
\begin{equation*}
{\#_N(\bigcap_{c_4 \in \mathcal{C}_4}\Bar{c_4})} \ge \begin{cases}
\frac{[Z(m_t+1)]^{\gamma \kappa}(1-\frac{2}{\Delta})^{\binom{\gamma}{2}\binom{\kappa}{2}}}{(\gamma !\kappa !)},&{\text{if}} \ I>II \\ 
{\frac{[Z(m_t+1)]^{\gamma \kappa}(1-\frac{\gamma \kappa-\gamma-\kappa+1}{4(\gamma \kappa-\gamma-\kappa)})^{\gamma\kappa}}{(\gamma !\kappa !)},}&{\text{otherwise.}} 
\end{cases}
\end{equation*} 
\end{corollary}

Above, we have obtained the lower bound of the solution space satisfying non-constructive LLL conditions. For constructive algorithms, we are also interested in how many solutions they can output from the solution space. We use the method from \cite{r18} that estimates the entropy of the MT distribution to derive a lower bound on the number of solutions the constructive algorithm can output.

\subsection{Diversity of Constructive Outputs via MT Entropy}
In this subsection, we use an entropy method (via Rényi entropy bounds for the MT output distribution) to obtain a computable lower bound on the number of feasible pairs $(\mathbf{P},\mathbf{L})$ that the algorithm can output.
Under the OO-CPO framework, the CPO operation is closely related to the MT algorithm (see Appendix F of \cite{r19}).
\begin{proposition}\label{p2}
In the OO-CPO framework, circulant power optimization (CPO) includes the Moser-Tardos (MT) algorithm as a special case.
\end{proposition}
\begin{IEEEproof}
This can be achieved by choosing the set of edges to be modified so that it corresponds to a specific harmful structure.
\end{IEEEproof}

In the sequel, we show that the MT algorithm can generate a provably large set of distinct feasible solutions. This observation suggests that CPO-based design procedures, such as those in~\cite{r12,r13,r14}, also have significant potential, since Proposition~\ref{p2} establishes that, within the OO-CPO framework, CPO subsumes the MT algorithm as a special case.

The MT distribution exhibits a high degree of randomness, similar to that of the original distribution. This is further quantified by the Rényi entropy of the MT distribution.

\begin{definition}
   Let $\mathcal{T}$ be a distribution on a finite set $\mathcal{S}$. We define the Rényi entropy with parameter $\alpha$ of $\mathcal{T}$ to be:
\begin{equation}
    H_{\alpha}({\mathcal{T}})=\frac{1}{1-\alpha}ln\sum_{t\in \mathcal{S}}P_{\mathcal{T}}(t)^{\alpha}
\end{equation} 
\end{definition}

 The entropy of any distribution is at most $ln(\mathcal{S})$, which is achieved by the uniform
 distribution.

We next quantify the \emph{output diversity} of MT Algorithm via the R\'enyi entropy of its output distribution.
Let $\Omega$ denote the product assignment space of the edge variables, and let $D_{\mathrm{MT}}$ be the distribution induced by the Moser--Tardos resampling procedure on $\Omega$.
Moreover, let $Ind (\mathcal A)$ denote the family of independent subsets of the bad-event set $\mathcal A$ in the corresponding dependency graph.
The following lemma provides a general lower bound on $H_\alpha(D_{\mathrm{MT}})$ for $\alpha>1$ in terms of the original product distribution and the LLL weights.

\begin{lemma}{(\cite{r18})}\label{lem_3}
Let $D_{MT}$ be the MT distribution.Then for $\alpha>1$,we have that
\begin{equation}
    H_{\alpha}(D_{MT})\ge  H_{\alpha}(\Omega)-\frac{\alpha}{\alpha-1}ln\sum_{I\in Ind(\mathcal{A})}\prod_{A\in I}\mu(A)
\end{equation} 
\end{lemma}

The term $\sum_{I\in Ind(\mathcal A)}\prod_{A\in I}\mu(A)$ can be interpreted as an independence polynomial (a hard-core partition function) on the dependency graph and is generally expensive to compute exactly.
To obtain a closed-form (but potentially looser) entropy bound that is easy to evaluate, we upper bound this term as follows.

\begin{lemma}{(\cite{r18})}\label{lem_4}
 We have that
\begin{equation}
    ln\sum_{I\in Ind(\mathcal{A})}\prod_{A\in I}\mu(A)\le \sum_{A\in \mathcal{A}}\mu(A)
\end{equation}

\begin{equation}
    H_{\alpha}(D_{MT})\ge  H_{\alpha}(\Omega)-\frac{\alpha}{\alpha-1}\sum_{A\in \mathcal{A}}\mu(A)
\end{equation}   
\end{lemma}
While Lemma~\ref{lem_4} yields a very simple bound, it may be overly conservative because it ignores the event--variable incidence structure.
To obtain a more informative yet still computable estimate, we further exploit $var(A)$ and distribute the weight $\mu(A)$ across the variables involved in each bad event.
This leads to the following product-form upper bound, parameterized by a per-event factor $l(A)$.

\begin{lemma}{(\cite{r18})}\label{lem_5}
For any bad event $A$, define

\begin{equation}
    l(A)=(1+\mu(A))^{\frac{1}{|var(A)|}}-1
\end{equation}
then,we have that
\begin{equation}
        \sum_{I\in Ind(\mathcal{A})}\prod_{A\in I}\mu(A)\le\prod_{i\in [n]}
        (1+\sum_{A\in \mathcal{A},i\in var(A)}l(A))
\end{equation}

\begin{equation}
    H_{\alpha}(D_{MT})\ge  H_{\alpha}(\Omega)-\frac{\alpha}{\alpha-1}ln\prod_{i\in [n]}
        (1+\sum_{A\in \mathcal{A},i\in var(A)}l(A))
\end{equation}    
\end{lemma}

With the above lower bounds on $H_\alpha(D_{\mathrm{MT}})$ in place, we now translate entropy into a diversity guarantee on the number of distinct feasible designs that MT Algorithm can output.
Specifically, for $\alpha>1$, the R\'enyi entropy of a distribution is upper bounded by the logarithm of its support size; hence a lower bound on $H_\alpha(D_{\mathrm{MT}})$ implies a lower bound on $|supp(D_{\mathrm{MT}})|$.
The next theorem formalizes this connection and, under a standard LLL-type weighting condition (involving the closed neighborhood $N^+(A_i)$), provides both an expected resampling bound and a computable lower bound on $\#_{\mathrm{MT}}(\cdot)$.

\begin{theorem}\label{the_3}
  For any QC-SC-LDPC code constructed with respect to a $\gamma \times \kappa$ fully-connected block code base graph,with no parallel edges, under the coupling pattern $\mathbf{a}=(a_0,a_1,...,a_{m_t})$ and corresponding probability distribution $\mathbf{p}=(p_0,p_1,...,p_{m_t})$. If $\{H_1,...,H_k\}$ are a series of avoidable harmful structures in the base graph with dependency graph $G$, and $\{P_{H_1},...,P_{H_k}\}$ are the probabilities that they are active after the random edge-spreading and lifting process,  if   
there is  a weighting function $\Tilde{\mu}: \mathcal{A}\rightarrow [0,\infty)$ satisfies
\begin{equation}
   \Tilde{\mu}(A_i)\ge P(A_i)\times \sum_{I\in Ind(N^+(A_i))}\prod_{A_j\in I}\mu(A_j)
\end{equation}
for any $i\in[n]$, then  the MT Algorithm  resamples an event $A\in \mathcal{A}$ at most an expected $\Tilde{\mu}(A_i)$ times before it finds such an evaluation.
The number of different edge-spreading and lifting assignments $(\mathbf{P},\mathbf{L})$ that the algorithm can output, denoted as $\#_{MT}(\bigcap_{i\in [k]}\Bar{H_i})$, satisfies the following condition:
\begin{align}
   \#_{MT}(\bigcap_{i\in [k]}\Bar{H_i})\ge exp(H_{\alpha}(D_{MT}))
\end{align}
\end{theorem}
\begin{IEEEproof}
see Appendix B of   ~\cite{r19}.
\end{IEEEproof}

\begin{corollary}
  For any QC-SC-LDPC code constructed with respect to a $\gamma \times \kappa$ fully-connected block code base graph,with no parallel edges, if
\begin{equation}\label{C3}
    \frac{2m^2+4m+3}{3(m+1)^3Z} \le \frac{(\Delta)^{\Delta}}{(\Delta+1)^{\Delta+1}}
\end{equation}
where $\Delta=(2\gamma-3)(2\kappa-3)$ , $Z$ is the lifting degree.   
then the number of different edge-spreading and lifting assignments $(\mathbf{P},\mathbf{L})$ that the algorithm can output, denoted as $\#_{MT}(\bigcap_{c_4\in \mathcal{C}_4}\Bar{c_4})$, satisfies the following condition:
\begin{align}\label{E19}
   \#_{MT}(\bigcap_{c_4\in \mathcal{C}_4}\Bar{c_4})\ge \frac{[Z(m_t+1)]^{\gamma \kappa}}{exp(\frac{\binom{\gamma}{2}\binom{\kappa}{2}}{4\gamma\kappa-6\gamma-6\kappa+8})}
\end{align}
Moreover, applying the alternative explicit specialization yields the following lower bound in terms of $l(A)$:

\begin{align}\label{E20}
    \#_{MT}(\bigcap_{c_4\in \mathcal{C}_4}\Bar{c_4})
   \ge \left(\frac{Z(m_t+1)}{\left(1+(\gamma-1)(\kappa-1)l(A)\right)}\right)^{\gamma\kappa}
\end{align}
\end{corollary}
\begin{IEEEproof}
The proof, the comparison of the two lower bounds, and the numerical verification are detailed in Appendices C, D, and E of \cite{r19}, respectively.
\end{IEEEproof}

\begin{corollary}
    The number $\#_{MT-N}(\bigcap_{i\in [k]}\Bar{H_i})$ of non-equivalent edge-spreading and lifting assignments $(\mathbf{P},\mathbf{L})$ satisfying Theorem~\ref{the_3}  that can be output by the algorithm is given by:
    \begin{align}
   \#_{MT-N}(\bigcap_{i\in [k]}\Bar{H_i})\ge \frac{exp(H_{\alpha}(D_{MT}))}{\gamma!\kappa!}
\end{align}
\end{corollary}

\begin{corollary}
    The number $\#_{MT-N}(\bigcap_{c_4\in \mathcal{C}_4}\Bar{c_4})$ of non-equivalent edge-spreading and lifting assignments satisfying Theorem~\ref{the_3} that can be output by the algorithm is given by:
\begin{align}
   \#_{MT-N}(\bigcap_{c_4\in \mathcal{C}_4}\Bar{c_4})\ge \frac{[Z(m_t+1)]^{\gamma \kappa}}{exp\left(\frac{\binom{\gamma}{2}\binom{\kappa}{2}}{4\gamma\kappa-6\gamma-6\kappa+8}\right)\gamma!\kappa!}
\end{align}
Analogous to Corollary 3,
\begin{align}
    \#_{MT-N}(\bigcap_{c_4\in \mathcal{C}_4}\Bar{c_4})
   \ge \frac{[Z(m_t+1)]^{\gamma\kappa}}{\gamma!\kappa![\left(1+(\gamma-1)(\kappa-1)l(A)\right)]^{\gamma\kappa}}
\end{align}
\end{corollary}

Let $n:=\gamma\kappa$ and $X:=Z(m_t+1)$. Corollary~3 provides an explicit \emph{diversity guarantee}
for the MT resampling procedure: under the stated LLL-type condition, the number of distinct feasible
edge-spreading and lifting assignments $(\mathbf{P},\mathbf{L})$ that MT algorithm can output while eliminating all $4$-cycle 
admits the closed-form lower bounds in \eqref{E19}--\eqref{E20}, whose dominant term is $X^{n}$ multiplied
by a degree-dependent penalty factor. Consequently,
\[
\#_{MT}\Bigl(\bigcap_{c_4\in\mathcal{C}_4}\overline{c_4}\Bigr)
= X^{n}\exp\!\bigl(-\Theta(n)\bigr)
= \exp\!\bigl(\Theta(n)\bigr),
\]
i.e., the MT output support is exponentially large in the number of edge variables.
Moreover, the product-form specialization in \eqref{E20} is strictly tighter than \eqref{E19}.
Corollary~5 further shows that, even after quotienting out row/column-permutation symmetries,
the number of \emph{non-equivalent} MT outputs remains provably large:
\[
\#_{MT\text{-}N}\Bigl(\bigcap_{c_4\in\mathcal{C}_4}\overline{c_4}\Bigr)
\ge \frac{1}{\gamma!\kappa!}\,
\#_{MT}\Bigl(\bigcap_{c_4\in\mathcal{C}_4}\overline{c_4}\Bigr),
\]
so the algorithm can still generate a large family of genuinely distinct structure-avoiding QC-SC-LDPC designs.

\section{Conclusion}
By casting structure-avoiding QC-SC-LDPC design as a CSP, we derived explicit lower bounds on the number of feasible edge-spreading and lifting assignments via a quantitative CLLL, allowing general non-uniform spreading and lifting distributions. We further obtained a computable diversity lower bound on the number of distinct solutions producible by the MT resampling algorithm using Rényi entropy bounds, with closed-form specializations for eliminating $4$-cycles. These bounds quantify the feasible design space and provide practical guidance for choosing memory/lifting parameters and sizing the remaining code-search space.

\section*{Acknowledgment}
This work is partially supported by the National Key R\&D Program of China, (2023YFA1009600).

\appendices

\section{Proof of the Theorem 1}\label{P1}
\setcounter{equation}{0}
\renewcommand{\theequation}{\thesection.\arabic{equation}}
Let $\Omega:=\big(\mathbf{a}\times\mathbb{Z}_Z\big)^{\gamma\kappa}=[(a_0,a_1,\ldots,a_{m_t})\times (0,1,\ldots,Z-1)]^{\gamma\kappa} $ be the assignment space of all edge-variables
$X_{i,j}=(\mathbf{P}(i,j),\mathbf{L}(i,j))$, and let
\begin{equation}
S:=\bigcap_{i\in[k]}\Bar{H_i}.
\end{equation}
be the set of satisfying assignments.
Under the assumed product distribution, for any assignment $\omega\in\Omega$,
\begin{align*}
\Pr_\Omega(\omega)&=\prod_{i\in[\gamma],\,j\in[\kappa]} 
\Pr[\mathbf{P}(i,j)=\omega^P_{i,j}]\Pr[\mathbf{L}(i,j)=\omega^L_{i,j}] \\
&\le (p_{\max}q_{\max})^{\gamma\kappa}.
\end{align*}

Hence,
\begin{equation}
\Pr_\Omega(S)=\sum_{\omega\in S}\Pr_\Omega(\omega)
\le \#(S)\cdot (p_{\max}q_{\max})^{\gamma\kappa}.
\end{equation}
which implies
\begin{equation}\label{eq:count_from_prob}
\#(S)\ge \left(\frac{1}{p_{\max}q_{\max}}\right)^{\gamma\kappa}\Pr_\Omega(S).
\end{equation}
By the quantitative clique-LLL bound used in \cite[Lemma 2]{r15} (under the stated condition on $\max_i P_{H_i}$),
\begin{equation}
\Pr_\Omega(S)\ge \prod_{v\in[m]}\Big(1-\sum_{i\in K_v}x_{i,v}\Big).
\end{equation}
Combining with \eqref{eq:count_from_prob} yields
\begin{equation}
\#(S)\ge \left(\frac{1}{p_{\max}q_{\max}}\right)^{\gamma\kappa}
\prod_{v\in[m]}\Big(1-\sum_{i\in K_v}x_{i,v}\Big).
\end{equation}
Finally, substituting into \cite[Theorem~1]{r15} the same choice of $\{x_{i,v}\}$ that yields the two explicit cases $I$ and $II$ gives the claimed bound.

\section{Proof of the Theorem 3}\label{P2}
\setcounter{equation}{0}
Since Lemmas \ref{lem_3}, \ref{lem_4}, and \ref{lem_5} are established in ~\cite{r18} based on the Quantitative Cluster Expansion Lovász Local Lemma (Lemma \ref{lem_7}), we adopt this specific formulation in the remainder of this paper for the sake of convenience. We note that these results can be readily generalized to more general forms of the Lovász Local Lemma; such extensions will be provided in an extended version of this work.In the context of the Quantitative Cluster Expansion Lovász Local Lemma, we can define the output distribution $D_{MT}$ of the MT algorithm analogously to the approach taken in \cite{r15}.

\begin{definition}
  The distribution of $\Omega$ conditioned on avoiding $\mathcal{A}$ is called the LLL-distribution. The distribution at the M-T algorithm termination is called the MT-distribution.  
\end{definition}

Let $D_{MT}$ be the MT output distribution on the (finite) set of assignments.
Let $S:=\mathrm{supp}(D_{MT})$ be its support, i.e., the set of assignments that Algorithm~1 can output with positive probability.
By definition, $\#_{MT}(\cap_{i\in[k]}\overline{H_i}) = |S|$.

We claim that for any distribution $D$ supported on a finite set $S$ and any $\alpha>1$,
\begin{equation}\label{eq:renyi_support}
H_\alpha(D)\ \le\ \ln |S|.
\end{equation}
Indeed, by Hölder (or power mean) inequality,
\begin{equation}
\Big(\sum_{s\in S} D(s)\Big)^\alpha \le |S|^{\alpha-1}\sum_{s\in S} D(s)^\alpha.
\end{equation}
Since $\sum_{s\in S} D(s)=1$, we obtain
\begin{equation}
\sum_{s\in S} D(s)^\alpha \ge |S|^{1-\alpha}.
\end{equation}
Taking logarithms and multiplying by $\frac{1}{1-\alpha}$ yields \eqref{eq:renyi_support}.

Applying \eqref{eq:renyi_support} to $D=D_{MT}$ gives
\begin{equation}
H_\alpha(D_{MT})\le \ln|S|
\quad\Longrightarrow\quad
|S|\ge \exp\big(H_\alpha(D_{MT})\big).
\end{equation}
Therefore,
\begin{equation}
\#_{MT}\Big(\bigcap_{i\in[k]}\overline{H_i}\Big)=|S|
\ge \exp\big(H_\alpha(D_{MT})\big),
\end{equation}
which completes the proof.

\section{Proof of the Corollary 3}\label{P3}
\setcounter{equation}{0}

Motivated by the reasons outlined in the proof of Theorem 3, and to facilitate the construction of a simple and concrete example, we focus on the Quantitative Lovász Local Lemma \ref{lem_6}. Note that for a given dependency graph $G_D$, the inclusion $\mathcal{I}_{a}(G_D)\subset\mathcal{I}_{CE}(G_D)$ holds ( Let $\mathcal{I}_{a}(G_D)$ and $\mathcal{I}_{CE}(G_D)$ be the interiors corresponding to Lemma \ref{lem_6} and Lemma \ref{lem_7}, respectively; for details, refer to Appendix F). Consequently, whenever the conditions of the Quantitative Lovász Local Lemma in Corollary 3 are satisfied, the conditions of Theorem 3 are automatically met. Following a line of reasoning similar to that in \cite{r22}, it follows that when (\ref{C3}) holds:

By Lemma~\ref{lem_4}, we have:
\begin{align}
H_\alpha(\Omega)
&= \frac{1}{1-\alpha}\ln\sum_{\omega\in\Omega}\Pr_\Omega(\omega)^\alpha \nonumber\\
&= \frac{1}{1-\alpha}\ln\Bigg(\sum_{(\ell,t)\in\{0,\ldots,m_t\}\times\mathbb{Z}_Z}(p_\ell q_t)^\alpha\Bigg)^{\gamma\kappa}\nonumber\\
&= \frac{\gamma\kappa}{1-\alpha}\ln\Bigg(\sum_{\ell=0}^{m_t}p_\ell^\alpha\Bigg)
   +\frac{\gamma\kappa}{1-\alpha}\ln\Bigg(\sum_{t\in\mathbb{Z}_Z}q_t^\alpha\Bigg). \label{eq:Halpha_Omega_pq}
\end{align}

\begin{align}
    & H_{\alpha}(D_{MT}) \ge  H_{\alpha}(\Omega)-\frac{\alpha}{\alpha-1}ln\sum_{I\in Ind(\mathcal{A})}\prod_{A\in I}\mu(A) \\   
     &\ge  \frac{1}{1-\alpha}ln\sum_{t\in \mathcal{S}}P_{\mathcal{T}}(t)^{\alpha}-\frac{\alpha}{\alpha-1}\sum_{A\in \mathcal{A}}\mu(A)\\
      &= \frac{1}{1-\alpha}ln\sum_{t\in \mathcal{S}}[Z(m_t+1)]^{-\gamma \kappa \alpha}-\frac{\alpha}{\alpha-1}\sum_{A\in \mathcal{A}}\frac{1}{\Delta-1}\\
      &= \frac{1}{1-\alpha}ln[Z(m_t+1)]^{(1-\alpha)\gamma\kappa}-\frac{\alpha}{\alpha-1}\frac{\binom{\gamma}{2}\binom{\kappa}{2}}{\Delta-1}\\
      &\overset{\alpha \to \infty}{=}ln[Z(m_t+1)]^{\gamma\kappa}-\frac{\binom{\gamma}{2}\binom{\kappa}{2}}{\Delta-1}.
\end{align}

\begin{align}
    \#_{MT}(\bigcap_{i\in [n]}\Bar{A_i})&\ge exp(H_{\alpha}(D_{MT}))\\
   &\ge \frac{[Z(m_t+1)]^{\gamma \kappa}}{exp(\frac{\binom{\gamma}{2}\binom{\kappa}{2}}{4\gamma\kappa-6\gamma-6\kappa+8})}.
\end{align}
Similarly, based on Lemma \ref{lem_5}, we have:
\begin{align}
    & H_{\alpha}(D_{MT}) \ge  H_{\alpha}(\Omega)-\frac{\alpha}{\alpha-1}ln\sum_{I\in Ind(\mathcal{A})}\prod_{A\in I}\mu(A) \\  
    &\ge  \frac{1}{1-\alpha}ln\sum_{t\in \mathcal{S}}P_{\mathcal{T}}(t)^{\alpha}-\frac{\alpha}{\alpha-1}ln\prod_{i\in [n]}(1+\sum_{\substack{A\in \mathcal{A}\\i\in var(A)}}l(A)).
\end{align}
For $\mathcal{H}=\mathcal{C}_4$, $|\mathrm{var}(A)|=8$
\begin{align}
       l(A)&=(1+\mu(A))^{\frac{1}{|var(A)|}}-1 =\left(\frac{\Delta}{\Delta-1}\right)^{\frac{1}{8}}-1\\
       &=\left(\frac{(2\gamma-3)(2\kappa-3)}{(2\gamma-3)(2\kappa-3)-1}\right)^{\frac{1}{8}}-1.
\end{align}

\begin{align}
 &ln \prod_{i\in [n]}(1+\sum_{\substack{A\in \mathcal{A}\\i\in var(A)}}l(A))=ln \prod_{i\in [n]}(1+W l(A))\\
 &=(\gamma\kappa)ln\left(1+(\gamma-1)(\kappa-1)l(A)\right).
\end{align} 
So
\begin{align}
    &H_{\alpha}(D_{MT}) \\
    &\ge  \frac{1}{1-\alpha}ln\sum_{t\in \mathcal{S}}P_{\mathcal{T}}(t)^{\alpha}-\frac{\alpha}{\alpha-1}ln\prod_{i\in [n]}(1+\sum_{\substack{A\in \mathcal{A}\\i\in var(A)}}l(A))   \\
    &=  \frac{1}{1-\alpha}ln\sum_{t\in \mathcal{S}}[Z(m_t+1)]^{-\gamma \kappa \alpha}\\
    &- \frac{\alpha}{\alpha-1} (\gamma\kappa)ln\left(1+(\gamma-1)(\kappa-1)l(A)\right) \\
    &\overset{\alpha \to \infty}{=}ln\left(\frac{Z(m_t+1)}{\left(1+(\gamma-1)(\kappa-1)l(A)\right)}\right)^{\gamma\kappa}.
\end{align}

\begin{align}
    \#_{MT}(\bigcap_{i\in [n]}\Bar{A_i})&\ge exp(H_{\alpha}(D_{MT}))\\
   &\ge \left(\frac{Z(m_t+1)}{\left(1+(\gamma-1)(\kappa-1)l(A)\right)}\right)^{\gamma\kappa}.
\end{align}

\section{Proof of the Comparison Inequality}\label{P4}
\setcounter{equation}{0}
\renewcommand{\theequation}{\thesection.\arabic{equation}}

Let $\gamma,\kappa\in\mathbb{Z}$ with $\gamma,\kappa>3$ and define
\begin{align}
X &:= Z(m_t+1), \label{eq:appA-X}\\
D &:= 4\gamma\kappa-6\gamma-6\kappa+8, \label{eq:appA-D}\\
l(A) &:=\left(\frac{(2\gamma-3)(2\kappa-3)}{(2\gamma-3)(2\kappa-3)-1}\right)^{\frac18}-1, \label{eq:appA-lA}\\
u &:= (\gamma-1)(\kappa-1)l(A), \label{eq:appA-u}\\
E &:= \frac{\binom{\gamma}{2}\binom{\kappa}{2}}{D}. \label{eq:appA-E}
\end{align}
Assume $X>0$. We prove the strict inequality
\begin{equation}
\left(\frac{X}{1+u}\right)^{\gamma\kappa} \;>\; \frac{X^{\gamma\kappa}}{\exp(E)} .
\label{eq:appA-goal}
\end{equation}

\begin{IEEEproof}
First note that
\begin{equation}
(2\gamma-3)(2\kappa-3)=4\gamma\kappa-6\gamma-6\kappa+9 = D+1.
\label{eq:appA-Dplus1}
\end{equation}
Since $\gamma,\kappa>3$, we have $\gamma,\kappa\ge 4$, hence $(2\gamma-3)(2\kappa-3)\ge 25$ and
\begin{equation}
D=(2\gamma-3)(2\kappa-3)-1 \ge 24 > 0.
\label{eq:appA-Dpos}
\end{equation}
Combining \eqref{eq:appA-lA} and \eqref{eq:appA-Dplus1} yields
\begin{equation}
l(A)=\left(1+\frac{1}{D}\right)^{\frac18}-1>0,
\label{eq:appA-lA-simplified}
\end{equation}
so $u>0$ and $1+u>1$.

Next, we upper bound $l(A)$. For $x>0$, the function $f(x)=(1+x)^{1/8}$ is strictly concave on $[0,\infty)$ since
\begin{equation}
f''(x)=\frac18\Bigl(\frac18-1\Bigr)(1+x)^{\frac18-2}<0.
\label{eq:appA-concave}
\end{equation}
Therefore, the tangent line at $x=0$ lies strictly above $f(x)$ for $x>0$, i.e.,
\begin{equation}
(1+x)^{\frac18} < 1+\frac{x}{8}, \qquad x>0.
\label{eq:appA-tangent}
\end{equation}
Applying \eqref{eq:appA-tangent} with $x=1/D$ gives
\begin{equation}
l(A)=\left(1+\frac{1}{D}\right)^{\frac18}-1 < \frac{1}{8D},
\label{eq:appA-lA-bound}
\end{equation}
and hence
\begin{equation}
u=(\gamma-1)(\kappa-1)l(A) < \frac{(\gamma-1)(\kappa-1)}{8D}.
\label{eq:appA-u-bound}
\end{equation}

We now compare $(1+u)^{\gamma\kappa}$ and $\exp(E)$. For $y>0$,
\begin{equation}
\ln(1+y) < y,
\label{eq:appA-log}
\end{equation}
because $g(y)=y-\ln(1+y)$ satisfies $g(0)=0$ and $g'(y)=y/(1+y)>0$ for $y>0$.
Using \eqref{eq:appA-log} with $y=u>0$ and exponentiating yields
\begin{equation}
(1+u)^{\gamma\kappa}
=\exp\!\bigl(\gamma\kappa\ln(1+u)\bigr)
<\exp(\gamma\kappa u).
\label{eq:appA-1u}
\end{equation}
Moreover, rewriting $E$ gives
\begin{equation}
E=\frac{\binom{\gamma}{2}\binom{\kappa}{2}}{D}
=\frac{\gamma\kappa(\gamma-1)(\kappa-1)}{4D}.
\label{eq:appA-E-rewrite}
\end{equation}
Combining \eqref{eq:appA-u-bound} and \eqref{eq:appA-E-rewrite} we obtain
\begin{equation}
\gamma\kappa u
<\gamma\kappa\cdot\frac{(\gamma-1)(\kappa-1)}{8D}
=\frac{E}{2}
<E,
\label{eq:appA-gku}
\end{equation}
and therefore
\begin{equation}
(1+u)^{\gamma\kappa} < \exp(\gamma\kappa u) < \exp(E).
\label{eq:appA-denom}
\end{equation}
Taking reciprocals in \eqref{eq:appA-denom} (all terms are positive) and multiplying by $X^{\gamma\kappa}>0$
proves \eqref{eq:appA-goal}.
\end{IEEEproof}

\section{Numerical Verification for $(\gamma,\kappa)=(3,5)$}
\setcounter{equation}{0}

For $(\gamma,\kappa)=(3,5)$ we have
\begin{align}
D &= 4\gamma\kappa-6\gamma-6\kappa+8
=4\cdot 15-18-30+8=20, \label{eq:appB-D}\\
l(A) &=\left(1+\frac{1}{D}\right)^{\frac18}-1
=\left(\frac{21}{20}\right)^{\frac18}-1\\
&\approx 0.00611740588712562, \label{eq:appB-lA}\\
u &= (\gamma-1)(\kappa-1)l(A)=8l(A)\\
&\approx 0.0489392470970049, \label{eq:appB-u}\\
E &= \frac{\binom{3}{2}\binom{5}{2}}{D}=\frac{3\cdot 10}{20}=1.5\\
\exp(E)&\approx 4.48168907033806. \label{eq:appB-E}
\end{align}
Hence
\begin{align}
(1+u)^{\gamma\kappa}&=(1+u)^{15}\approx (1.0489392470970049)^{15}\\
&\approx 2.04764671546561\\
&< \exp(E)\approx 4.48168907033806.
\label{eq:appB-check}
\end{align}
For instance, setting $X=1$ gives
\begin{align}
\left(\frac{1}{1+u}\right)^{15}&\approx 0.488365494129007\\
&>\exp(-E)=e^{-1.5}\approx 0.223130160148430,
\label{eq:appB-check2}
\end{align}
which numerically confirms the comparison.

\section{Auxiliary Lemmas}\label{Aux}
\setcounter{equation}{0}

The Lovász local lemma (LLL), introduced by Erdős and Lovász in 1975, has become one of the most important probabilistic methods. LLL provides sufficient conditions under which a set of undesirable events $\mathcal{A}$  in a probability space $\Omega$ can be simultaneously avoided, meaning the conditions for  $\Pr(\cap_{A\in \mathcal{A}}\Bar{A})> 0$ to hold. The relationships between these bad events are depicted by an undirected graph called the \textit{dependency graph} $G_D=([n],E)$, where each bad event corresponds to a vertex in the dependency graph. An event $A_i$ is independent of $\{A_j:j\neq i,j\notin \mathcal{N}(i)\}$, where $\mathcal{N}(i)$ stands for the neighborhood of $A_i$ in $G$.

Note that in the variable-generated system model, the event-variable graph $G_B=([n],[m'],E)$ not only captures the dependency relationships between events but also the strength of these dependencies, which the dependency graph $G_D=([n],E)$ cannot capture. Therefore, the event-variable graph contains more information than the dependency graph. Given the bipartite graph $G_B=([n],[m'],E)$, let $E'=\{(v_1,v_2) : v_1,v_2 \in [n], \mathcal{N}(v_1) \cap \mathcal{N}(v_2) \neq \emptyset\}$. Define the dependency graph corresponding to the event-variable graph $G_B=([n],[m'],E)$ as $G_D(G_B)=([n],E')$. 

We use $\mathcal{A} \sim (G_D, \mathbf{p})$ to denote that (i) $G_D$ is a dependency graph of $\mathcal{A}$ and (ii) the probability vector of $\mathcal{A}$ is $\mathbf{p}$. Given a dependency graph $G_D$, define the \textit{abstract interior} $\mathcal{I}_{a}(G_D)$ to be the set consisting of all vectors $\mathbf{p}$ such that $\Pr\left(\bigcap_{A \in \mathcal{A}} A \right) > 0$ for any $\mathcal{A} \sim (G_D, \mathbf{p})$. We abbreviate $\mathcal{I}{a}(G_D(G_B))$ as $\mathcal{I}{a}(G_B)$. Similarly, We use $\mathcal{A} \sim (G_B, \mathbf{p})$ to denote that (i) $H_E$ is a event-variable graph of $\mathcal{A}$ and (ii) the probability vector of $\mathcal{A}$ is $\mathbf{p}$. In this appendix, we reuse the symbol $\mathbf{p}$ to denote a set of probability vectors, distinct from its usage in the main text.

LLL offers a powerful method to demonstrate the existence of complex combinatorial objects that satisfy a given set of requirements. The initial result for abstract LLL was established by Erdős and Lovász, and the first asymmetric version (Lemma \ref{lem_6}) was introduced later. 
In the following, we will introduce a series of LLL variants that offer trade-offs between bounds and complexity.

\begin{lemma}\label{lem_6}{(Quantitative Lovász Local Lemma)} Given a dependency graph $G=([n],E)$, and a probability vector $\mathbf{p}=(P(A_1),\ldots,P(A_n)) \in (0,1)^n$, if there exist real numbers $x_1,\ldots,x_n\in(0,1)$ such that
\begin{equation}
    P(A_i) \le x_i\prod_{j\in \mathcal{N}(i)}(1-x_j)
\end{equation}
for any $i\in[n]$, then 
\begin{equation}
\Pr(\cap_{A\in \mathcal{A}}\Bar{A})>\prod_{j\in [n]}(1-x_j)
\end{equation} 
\end{lemma}

By utilizing the relationship between the Lovász Local Lemma and the partition function of the hard-core lattice gas on graphs, an improved version of the Lovász Local Lemma was obtained in~\cite{r23}.

\begin{lemma}\label{lem_7}{(Quantitative Cluster Expansion Lovász Local Lemma)} Given a dependency graph $G=([n],E)$, and a probability vector $\mathbf{p}=(P(A_1),\ldots,P(A_n)) \in (0,1)^n$. Suppose that a weighting function $\Tilde{\mu}: \mathcal{A}\rightarrow [0,\infty)$ satisfies
\begin{equation}
   \Tilde{\mu}(A_i)\ge P(A_i)\times \sum_{I\in Ind(\mathcal{N}^+(A_i))}\prod_{A_j\in I}\mu(A_j)
\end{equation}
for any $i\in[n]$, where $Ind(G)$ is the collection of independent sets of $G$, then 
\begin{equation}
\Pr(\cap_{A\in \mathcal{A}}\Bar{A})>\prod_{j\in [n]}(1-P(A_j))^{\varphi_{\Tilde{\mu}}(A_j)}
\end{equation} 
with 
\begin{equation}
\varphi_{\Tilde{\mu}}(A_j) = \sum_{I\in Ind(\mathcal{N}(A_j))}\prod_{A_k\in I}\mu(A_k)
\end{equation} 
\end{lemma}

The original LLL (Lemma 1) only uses minimalistic local structural information. In~\cite{r24}, a hierarchy of LLLs is presented, which are increasingly complex and utilize an increasing amount of local information in a non-trivial way. In the limit, they yield Shearer's condition, which uses all the global information regarding the structure of the dependency graph.

\begin{lemma}\label{lem_8}{(Quantitative Clique Lovász Local Lemma)}
Let $\mathcal{A}=\{A_1,A_2,\ldots,A_n\}$ be a set of events with dependency graph $G$ and let $\mathcal{K}=\{K_1,K_2,\ldots,K_m\}$ be a set of cliques in $G$ covering all the edges (not necessarily disjointly). If there exist a set of vectors $\{\textbf{x}_1,\textbf{x}_2,\ldots,\textbf{x}_m\}$ from $(0,1)^n$ such that following conditions are satisfied:
\begin{enumerate}
    \item $\forall v \in [m]: \sum_{i\in K_v}x_{i,v}<1$
    \item $\forall i \in [n],\forall v$ such that $i\in K_v$:
    \begin{equation}
        P(A_i)\le x_{i,v}\prod_{u\neq v:i\in K_u}(1-\sum_{j\in K_u \backslash \{i\}}x_{j,u})
    \end{equation}
\end{enumerate}
then:
\begin{enumerate}
    \item $\Pr(\bigcap_{i\in [n]}\Bar{A_i})\ge \prod_{v\in [m]}(1-\sum_{i\in K_v}x_{i,v})>0$
    \item In the variable framework, the running time of the algorithm of Moser and Tardos is at most:
    \begin{equation}
        \sum_{i\in [n]}min_{v:i\in K_v}\frac{x_{i,v}}{1-\sum_{j\in K_v}x_{j,v}}
    \end{equation}
\end{enumerate}
\end{lemma}

\begin{figure}[!ht]
		\label{f2}
		\renewcommand{\algorithmicrequire}{\textbf{Input:}}
		\renewcommand{\algorithmicensure}{\textbf{Output:}}
		%\removelatexerror
		\begin{algorithm}[H]
			\caption{The Moser-Tardos (MT) Algorithm }
			\begin{algorithmic}[1]\label{alg_1}
				\REQUIRE  
                          Sample space: $\mathbf{a}=(a_0,a_1,\ldots,a_{m_t})$.\\
                          Independent r.v. taking values in $\mathbf{a}$: $\chi=\{X_1,...,X_{\gamma\kappa}\}$.\\
                          Set of (ordered) events: $\mathcal{H}=\{H_1,...,H_k\}.$\\
                          Probability distribution: $\mathbf{p}=(p_0,p_1,\ldots,p_{m_t})$.
				\ENSURE   Assignment $\alpha =(r_1,...,r_{\gamma\kappa})\in \mathbf{a}^{\gamma\kappa}$ to  r.v.   
                          $\chi$  s.t. $\cap_{H_i \in \mathcal{H}}\Bar{H_i}=TRUE$.
				\STATE Sample the variables $X_i,~i\in[\gamma\kappa]$, and let $\alpha$ be the resulting assignment.
				\WHILE {there exists a bad event in $\mathcal{H}$ that occurs under the current assignment, let $H_j$ be the least indexed such event}
				\STATE RESAMPLE($H_j$)
				\ENDWHILE
                \STATE Output current assignment $\alpha$.
			\end{algorithmic}
            RESAMPLE($H_j$)
            \begin{algorithmic}[1]
                    \STATE Resample the variables in sc($H_j$).
                    \WHILE {there is a least indexed bad event $H_l$, such that sc($H_j$) $\cap$ sc($H_l$) $\neq \emptyset$, occurring under the current assignment,}
                    \STATE RESAMPLE($H_l$)
                    \ENDWHILE
            \end{algorithmic}
		\end{algorithm}
	\end{figure}

\begin{comment}
   \section{CONCLUSION}
\label{con}
This paper uses probabilistic combinatorial methods to construct QC-SC-LDPC codes. To improve error-floor performance, harmful combinational structures must be removed. Using the Lovász Local Lemma (LLL), we derive upper bounds on memory and lifting degree, and propose an algorithm with guaranteed theoretical performance. We also analyze interactions between different harmful structures and provide explicit bounds. 
\end{comment}

%\IEEEtriggeratref{14}

\bibliographystyle{IEEEtran}

\bibliography{IEEEabrv,ref.bib}

\end{document}